\documentclass[journal,transmag]{IEEEtran}

\usepackage{graphicx}
\usepackage{amsmath}
\usepackage{amssymb}
\usepackage{cite}
\usepackage{xcolor}
\usepackage{epstopdf}

\begin{document}

\title{Deep Learning-Based Decoding of Linear Block Codes for Spin-Torque Transfer Magnetic Random Access Memory (STT-MRAM)}


\author{\IEEEauthorblockN{Xingwei Zhong\IEEEauthorrefmark{1},
Kui Cai\IEEEauthorrefmark{1},~\IEEEmembership{Senior Member,~IEEE},
Zhen Mei\IEEEauthorrefmark{2,1}, and Tony Q. S. Quek\IEEEauthorrefmark{1},~\IEEEmembership{Fellow,~IEEE}}
\IEEEauthorblockA{\IEEEauthorrefmark{1} Singapore University of Technology and Design (SUTD), Singapore, 487372}
\IEEEauthorblockA{\IEEEauthorrefmark{2} Huawei Technologies Co. Ltd., Shenzhen, P. R. China, 518129}
}




\IEEEtitleabstractindextext{
\begin{abstract}
Thanks to its superior features of fast read/write speed and low power consumption, spin-torque transfer magnetic random access memory (STT-MRAM) has become a promising non-volatile memory (NVM) technology that is suitable for many applications. However, the reliability of STT-MRAM is seriously affected by the variation of the memory fabrication process and the working temperature, and the later will lead to an unknown offset of the channel. Hence, there is a pressing need to develop more effective error correction coding techniques to tackle these imperfections and improve the reliability of STT-MRAM. In this work, we propose, for the first time, the application of deep-learning (DL) based algorithms and techniques to improve the decoding performance of linear block codes with short codeword lengths for STT-MRAM. We formulate the belief propagation (BP) decoding of linear block code as a neural network (NN), and propose a novel neural normalized-offset reliability-based min-sum (NNORB-MS) decoding algorithm. We successfully apply our proposed decoding algorithm to the STT-MRAM channel through channel symmetrization to overcome the channel asymmetry. We also propose an NN {based} soft information generation method (SIGM) to take into account the unknown offset of the channel. Simulation results demonstrate that our proposed NNORB-MS decoding algorithm can achieve significant performance gain over both the hard-decision decoding (HDD) and the regular  {reliability-based min-sum (RB-MS)} decoding algorithm, for cases without and with the unknown channel offset. Moreover, the decoder structure and time complexity of the NNORB-MS algorithm remain similar to those of the regular RB-MS algorithm.

\end{abstract}

}

\maketitle

\IEEEdisplaynontitleabstractindextext

%
\IEEEpeerreviewmaketitle

\section{Introduction}

Spin-torque transfer magnetic random access memory (STT-MRAM) is an advanced non-volatile memory (NVM) technology with high endurance, fast read/write access time, and low switching energy \cite{yu2016emerging}. However, process variation and thermal fluctuation lead to widened distributions of the low and high resistance states of STT-MRAM and their overlapping, and hence the memory sensing errors \cite{cai2013channel}. Furthermore, the increase of temperature will also cause the high resistance to decrease, while the low resistance remains unchanged \cite{wu2016temperature}, leading to a channel offset unknown during the readback process and a degradation of the error rate performance.


To be compatible with the fast read/write speed of STT-MRAM, error correction codes (ECCs) with short codeword lengths of less than a few hundred bits are employed. A (71,64) Hamming code with single-error correction capabilities is  used for Everspin's 16Mb MRAM \cite{mram}. To further improve the memory storage density, multiple-error-correcting Bose-Chaudhuri-Hoquenghem (BCH) codes are proposed \cite{del2014improving}. However, so far both the Hamming codes and BCH codes are decoded using hard-decision decoding (HDD)  {through the standard bounded distance decoders \cite{richardson2008modern}. The regular soft-decision decoding (SDD) decoders for these codes are the standard belief propagation (BP) decoders \cite{richardson2008modern}, such as the sum and product algorithm (SPA) based decoder, the min-sum (MS) decoder, and the reliability-based min-sum (RB-MS) decoder \cite{chen2011comparisons}, which are performed over the Tanner graph of the codes. They provide little to no performance gain over the HDD decoders for short codes.}

To tackle the variation of the STT-MRAM channel such as the unknown channel  {offset caused by the temperature change}, the state of the art approach is to introduce reference cells with known data to the memory array to estimate the channel dynamically \cite{na2014reference}. However, a frequent insertion of reference cells will lead to a high redundancy and hence degrade the information storage efficiency. Furthermore, the reference cell itself may also suffer from the non-uniformity issue caused by fabrication imperfection which may affect the accuracy of channel estimation.


On the other hand, in recent years, the deep learning (DL) techniques have achieved great success in many areas, such as speech recognition and image processing \cite{egmont2002image}. It has also been shown that DL methods \cite{nachmani2016learning,lugosch2017neural} can be used to improve the performance of standard SDD  {decoders}, for high-density and short algebraic codes. However, in these works, the channel must be symmetric to ensure that the decoding error rate is independent of the transmitted codeword, otherwise the neural network (NN)-based decoder cannot be trained using a single codeword \cite{nachmani2016learning,lugosch2017neural}. Besides that, to the best of our knowledge, no work has been reported on DL-based channel decoding for STT-MRAM.

In this work, we propose novel DL-based algorithms to improve the decoding performance of linear block codes with short codeword lengths for STT-MRAM. A prerequisite for NN-based decoding is that the channel must be symmetric while the STT-MRAM channel is asymmetric by nature. Therefore, we adopt an independent and identically distributed (i.i.d.) channel adapter \cite{hou2003capacity} to `symmetrize' the STT-MRAM channel, and hence only the all-zero codeword is needed during the training of the NN decoder. Another obstacle for applying SDD is that it is difficult to obtain the channel soft information in the presence of  {the} unknown channel offset. In this work, we present a novel NN-based  {soft information generation method (SIGM)}, which can estimate  {the} channel soft information by using a few bits of uniform quantization. It only needs to be activated when the ECC decoder fails, and hence avoids a significant increase of the read latency and power consumption. We further formulate the BP decoding as a NN, and propose a novel neural normalized-offset reliability-based min-sum (NNORB-MS) decoding algorithm. We demonstrate that it can significantly outperform both the HDD and the regular  {RB-MS} algorithm \cite{chen2011comparisons}, for cases without and with the unknown channel offset. Moreover, the decoder structure and time complexity of the NNORB-MS algorithm remain similar to those of the regular RB-MS algorithm.



The rest of the paper is organized as follows. In Section II, we describe the STT-MRAM channel model and its symmetrization. In Section III, we present a novel NN-based SIGM which can estimate the channel soft information without the prior knowledge of the channel. A NNORB-MS algorithm for the STT-MRAM channel is proposed in Section IV and the simulation results are illustrated in Section V. Finally, Section VI concludes the paper.


\section{System Modeling}
\subsection{STT-MRAM Channel Model}
An STT-MRAM cell consists of a magnetic tunneling junction (MTJ) as the data storage element and an nMOS transistor as the access control device. The MTJ has a tunneling oxide layer sandwiched between a free layer and a reference layer. While the magnetization direction of the free layer can be switched by passing write currents of different directions through the MTJ, that of the reference layer is always fixed. When the magnetization direction of the free layer is different from that of the reference layer, the MTJ has a high resistance, which can be used to denote a binary input bit `1'. Otherwise if the magnetization direction of the free layer is the same as that of the reference layer, the MTJ has a low resistance, which can be used to denote an input bit `0'.

\begin{figure}[t]
\centering
\includegraphics[height=0.3\columnwidth]{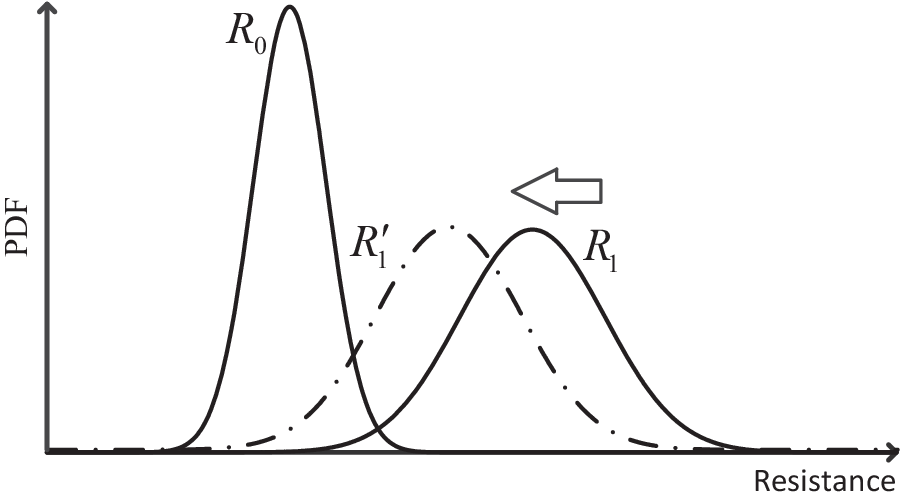}
\caption{Distributions of resistances of STT-MRAM: $R_{0}$ indicates the low resistance, $R_{1}$ represents the original high resistance, and $R_{1}^{'}$ is the shifted high resistance caused by temperature increase.}
\label{pdf_model}
\end{figure}

It has been widely observed that the reliability of data stored in the STT-MRAM cell is affected by the process imperfection, which causes variations of both the MTJ geometry and the nMOS transistor size, leading to widened distributions of  {the two} resistance states and their overlapping \cite{cai2013channel,cai2017cascaded}. Moreover, the working temperature  {also affects} the resistance distributions of STT-MRAM. It has been found that with the increase of the temperature, the high resistance $R_{1}$ decreases, while the low resistance $R_{0}$ remains almost the same \cite{wu2016temperature}. The probability density functions (PDFs) of the resistances for STT-MRAM are shown in Fig. \ref{pdf_model}. Based on the stochastic characteristics of the two resistance states  {and by following \cite{mei2019neural}}, the vector of resistances read back from the STT-MRAM cell can be modeled as
\begin{equation} \label{model}
\textbf{\textit{y}} = \textbf{\textit{r}} + \textbf{\textit{n}} + \textbf{\textit{b}},
\end{equation}
where $\textbf{\textit{r}} =\lbrace r_{1},\cdots,r_{N} \rbrace$ are the nominal resistance values of the memory cells that store {the} user data bits $\textbf{\textit{x}} =\lbrace x_{1},\cdots,x_{N} \rbrace$, $x_{l}\in\lbrace 0,1 \rbrace$, $l \in {1,\cdots,N}$, and $N$ is the length of the linear block code. Thus, $r_{l} = \mu_{0}$ for $x_{l} = 0$, and $r_{l} = \mu_{1}$ for $x_{l} = 1$. The noise vector $\textbf{\textit{n}} =\lbrace n_{1},\ldots,n_{N} \rbrace$ in (1) accounts for the variations of the resistances $R_{0}$ and $R_{1}$ caused by process imperfection, with $n_{l}$ being i.i.d. with a zero-mean and a variance of $\sigma_{i}^{2}$, $i = 0,1$. Note that the distribution of $n_{l}$ can be non-Gaussian. Furthermore, $\textbf{\textit{b}} =\lbrace b_{1},\ldots,b_{N} \rbrace$ denotes the unknown offset caused by the temperature change  {. The offset $b_{l}$ only occurs with the high resistance state when $x_{l} = 1$, and} $b_{l} = 0$ for $x_{l} = 0$.  {Moreover, as STT-MRAM suffers from the process variation induced variations of the access transistor size and the magnetic tunneling junction geometry, the influence of temperature on each memory cell is random. Hence we assume that the offset $b_{l}$ associated with $x_{l} = 1$ is i.i.d, and it follows a Gaussian distribution $N(\mu_{b},\sigma^2_{b})$ with a mean of $\mu_{b}$ and standard deviation of $\sigma_{b}$.}

 {We remark that although there are papers in the literature investigated the thermal impact on STT-MRAM theoretically or experimentally \cite{wu2016temperature},  due to the complication of memory physics, it is difficult to predict the offset $b_l$ over different temperatures accurately. Hence the actual amount of the offset $b_l$ is unknown to the channel detector and decoder. The channel model of (1) is only used to carry out channel simulations and to generate data for training and testing the NN based detector and decoder, since it is difficult to obtain the practical data of STT-MRAM.
}

\subsection{Symmetrization of the STT-MRAM Channel}
The error rate performance of  {the standard SDD} decoders can be improved by  {using} the recently proposed NN-based decoders \cite{nachmani2016learning,lugosch2017neural}. The corresponding decoding algorithms satisfy the message passing symmetry conditions, and their error rates are independent of the transmitted codeword when transmitting over a binary memoryless symmetric channel \cite{richardson2008modern}. As a result, it is sufficient to train the NN decoder using a single codeword ({\it e.g.} the all-zero codeword), thus avoiding the curse of dimensionality for NN-based decoding \cite{nachmani2016learning,lugosch2017neural}.

A binary-input channel is symmetric if  {$p(Y = y|X = 0) = p(Y = -y|X = 1)
$}, with $Y$ and $X$ being the output and input of the channel, respectively  {\cite{hou2003capacity}}. The STT-MRAM channel is obviously asymmetric. In this work, we adopt an i.i.d. channel adapter \cite{hou2003capacity} to force the STT-MRAM channel to be symmetric. It mainly involves the following steps. In the first step, generate binary symbols $p_{l}$ from an i.i.d. source with equiprobable distribution. After that, these symbols are mod-2 added with the channel input bits $x_{l}$ and transmitted over the channel. At the receiver, adopt a sign adjuster which performs the operation $v_{l} = u_{l} * (1 - 2 * p_{l})$, with $u_{l}$ being the log-likelihood ratio (LLR) of the channel output bits, $v_{l}$ being the input of the  {SDD} decoder, respectively.  {The generation of channel soft information $u_{l}$ will be discussed} in the next section. It can be proved that the newly augmented channel with input bit $x_l$ and output bit $v_{l}$ is a binary-input symmetric channel \cite{hou2003capacity}, and hence the original STT-MRAM channel is transformed into a `symmetrized channel' .

\section{A NN-based Soft Information Generation Method}
The accuracy of the channel soft information sent to the decoder will significantly affect the decoding performance. For the STT-MRAM channel with unknown offset, it is difficult to derive  {the} channel soft information directly. Although signals read back from reference cells can be used to estimate the mean of the offset $b_{l}$, the accuracy of soft information will still degrade since the variance of $b_{l}$ cannot be taken into account. In this work, we propose a novel NN-based SIGM, which can accurately estimate the channel soft information of the STT-MRAM channel in the presence of unknown channel offset.


Since more errors will occur near the overlapping region of  {the PDFs of the two resistance states, quantization should be concentrated around this region. In} the presence of unknown channel offset, our key idea is to first use the outputs of an NN-based  {detector developed by \cite{mei2019neural}} to find an appropriate hard detection threshold of the  {channel, which indicates the center of the overlapping region of the two resistance states. Optimum boundaries of a multiple-bit quantizer} for SDD can then be determined subsequently.

To determine an appropriate hard detection threshold for the STT-MRAM channel with unknown channel offset, we follow the approach presented in \cite{mei2019neural}.  {That is, we first develop an recurrent neural network (RNN) based channel detector, which can achieve better error rate performance than the multilayer perceptron (MLP) detector with smaller size of training data. The details of the RNN structure can be found in \cite{mei2019neural}.} The inputs to the NN are the resistance
values $\textbf{\textit{y}}$ read back from the memory cells, and the outputs of the NN are the soft
estimates $\tilde{\textbf{\textit{x}}}$ of the binary bit sequence \textbf{\textit{x}} stored in the memory array.
We can obtain the hard estimation $\bar{\textbf{\textit{x}}}$ of $\textbf{\textit{x}}$ based on the following hard decision rule: if $\tilde{x}_l>0.5$, $\bar{x}_l=1$; otherwise  {$\bar{x}_l=0$}. Moreover, with an assumed detection threshold $R_t$, we can get  {the hard estimation $\hat{\textbf{\textit{x}}}$} for a given $\textbf{\textit{y}}$. Therefore, an adjusted detection threshold $\tilde{R_t}$ can be obtained by searching for a threshold $R_t$ that minimizes the Hamming distance between $\bar{\textbf{\textit{x}}}$ and  {$\hat{\textbf{\textit{x}}}$}.

\begin{figure}[t]
\centering
\includegraphics[height=0.35\columnwidth]{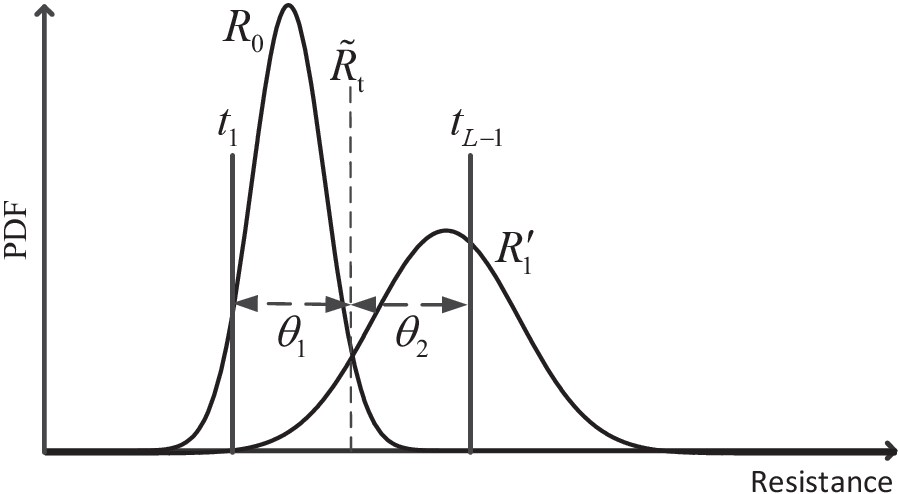}
\caption{Proposed quantization of the STT-MRAM channel.}
\label{intervals_region}
\end{figure}

Once we obtain the adjusted detection threshold $\tilde{R_t}$  {using the RNN detector presented by \cite{mei2019neural}, we propose the following scheme to further design a multiple-bit quantizer that can generate the channel log likelihood ratios (LLRs) for SDD.} As shown by Fig. \ref{intervals_region}, with $t_{0}=-\infty$ and $t_{L}=+\infty$, $t_{1}=\tilde{R_t}-\theta_1$, and $t_{L-1}=\tilde{R_t}+\theta_2$, we define a quantizer that uniformly divides the quantization range $[ t_{1},t_{L-1} ]$ into $2^{q}-2$ quantization intervals, with $L=2^{q}$ being the number of quantization levels, $q$ being the total number of the quantization bits, respectively. Next, to facilitate the usage of reliability-based soft-decision decoders, we set the LLR of the channel received bit to be the integer index of the quantization region that the corresponding channel readback signal falls in. Furthermore, if the quantization region is closer to $\tilde{R_t}$, it is more difficult to differentiate between `0' and `1'; while as the quantization region is getting away from $\tilde{R_t}$, we are more likely to distinguish between `0' and 1'. Based on these observations, the quantization index of the $s$-th quantization interval $[ t_{s},t_{s+1} ]$ is set to be $I_s=s-q$, for $s = 0,1,\cdots, L-1$.

In this way, the LLRs of the channel received bits can be determined for given values of $t_1$ and $t_{L-1}$.  {To determine} the optimum values of $t_1$ and $t_{L-1}$,  {we adopt a computer exhaustive search approach to tune $\theta_1$ and $\theta_2$ such that the corresponding LLRs generated lead to the best error rate performance of the RB-MS decoder.} Simulation results show that by using only a $q=3$ bits quantizer, we can achieve decoding performance similar to that with the full channel soft information ({\it i.e.} $q=\infty$). We have also explored the non-uniform quantizer within the quantization range $[ t_{1},t_{L-1} ]$, and found that its performance gain over the uniform quantizer is negligible. Note that the above proposed NN-based  {SIGM} only needs to be activated when the ECC decoder fails and can be terminated once the updated quantizer is determined, and hence it avoids a significant increase of the read latency and power consumption compared to the case of  {applying the NN-based SIGM} for every input data block.  {Note also that the LLRs generated by the proposed NN-based SIGM can be used by any SDD decoders, including the RB-MS and the NNORB-MS decoders that will be discussed in the next section. }

\section{NNORB-MS Algorithm for STT-MRAM}


The RB-MS algorithm \cite{chen2011comparisons} takes the integer index of the quantization interval that the received signal falls in as the channel LLR for decoding. It requires only integer and logic operations, and hence has a lower computational complexity and can meet the low latency requirement of STT-MRAM. In this work, we first set-up a trellis that corresponds to an NN to represent the regular RB-MS decoding with $T$ full iterations instead of using the conventional Tanner graph. For an ECC of length $N$ bits, the input layer of the trellis is a vector of size $N$, {consisting of the LLRs} of the channel output bits generated by the NN-based SIGM described in the previous section. The nodes in the hidden layer correspond to edges in the Tanner graph, and the associated processing elements carry out the check nodes (CNs) and variable nodes (VNs) operations for decoding. The output layer has $N$ processing elements that output the decoded codeword.

We further propose a novel NNORB-MS algorithm to improve the decoding performance of ECCs with short codeword lengths for STT-MRAM. {In particular, we introduce two decoding parameters to the RB-MS algorithm, namely the offset $\beta_{c,k}$ for edge connecting the $c$-th CN to the $k$-th VN and the normalization factor $\delta_{k}$ for the $k$-th VN. These parameters can correct the amplified magnitude of the message introduced by the MS algorithm, and also compensate for the small cycles of the short linear block codes.  We propose to set these two key decoding parameters as the learnable weights of the NN. Through the training of the NN, we can obtain the optimum values of $\beta_{c,k}$ and $\delta_{k}$ that lead to the best error rate performance of decoding. Except for the introduction of $\beta_{c,k}$ and $\delta_{k}$, the corresponding operations of the NNORB-MS algorithm at the CN and VN are similar to those of the RB-MS algorithm, which are summarized as follows.}

(a) Check-node update:
\begin{align} \nonumber
&\varepsilon_{c,k}^{(t)} = \\
&\prod_{k^{'} \in N(c)\setminus k}sign(\lambda_{k^{'},c}^{(t-1)}) &\cdot ReLU \left( \min_{k^{'} \in N(c)\setminus k} (\vert \lambda_{k^{'},c}^{(t-1)} - \beta_{c ,k} \vert) \right),
\end{align}
where $\varepsilon_{c,k}^{(t)}$ is the message from the $c$-th CN to the $k$-th VN at the $t$-th iteration,  {$N(c)\setminus k$ is the set of VNs connected to CNs excluding the $k$-th VN}, and $\beta_{c,k}$ denotes the offset to be learned from the NN.  {Here, $ReLU(.)$ is a ``rectifier” activation function \cite{lugosch2017neural} that can prevent the subtraction in (2) from flipping the sign of the message. In (2), $\lambda_{k^{'},c}^{(t-1)}$ is the} input message from the $k^{'}$-th VN to the $c$-th CN for the previous $(t-1)$-th  {iteration, given by}:
\begin{equation} \label{Previous-btc1}
\lambda_{k^{'},c}^{(t-1)} = \xi_{k^{'}}^{(t-1)} - \lambda_{c,k^{'}}^{(t-1)},
\end{equation}
where $\xi_{k^{'}}^{(t-1)}$ is the {\it a posteriori} integer information for iteration $(t-1)$ and  {$\lambda_{c,k^{'}}^{(t-1)}$ denotes the input message from the $c$-th CN to $k^{'}$-th VN for the $(t-1)$-th iteration}.

(b) Variable-Node update:
\begin{equation} \label{Bit-Node1}
\xi_{k}^{(t)} = \left[ \lambda_{k} + \delta_{k} \times \sum_{c^{'} \in M(k)} \varepsilon_{c^{'},k}^{(t)} \right],
\end{equation}
where  {$\xi_{k}^{(t)}$ represents the message from the $k$-th VN for iteration $(t)$, $\lambda_{k}$ is a priori information of bit $k$}, $\delta_{k}$ denotes the normalization factor to be learned from the NN, $M(k)$ is the set of CNs connected to VNs, and $\left[ \cdot \right]$ is the operation of rounding the number inside the bracket to the closest integer.

(c)  {Generation of soft version of the  hard decision:}
\begin{equation} \label{soft-hard decision}
\sigma(\xi_{k}^{(t)}) = \dfrac{1}{1+e^{-\xi_{k}^{(t)}}},
\end{equation}
where $\sigma(\cdot)$ is a sigmoid function that enables the cross-entropy loss function and multiloss function to be used in the training process of the NN. Finally, after the complete NN  {processing}, the decoded bits can be derived as  {the} simple hard decisions.

In our experiments, we adopt an RNN structure,  {as we found that} it can reduce the total number of NN parameters. By applying the deep unfolding technique \cite{hershey2014deep}, the RNN is chosen to have 10 hidden layers corresponding to 5 full decoding iterations. The NN is trained according to the Adam update rule \cite{nachmani2016learning,lugosch2017neural} with a learning rate of 0.01 and 10000 minibatches of 100 codewords each. Due to the symmetrization of the STT-MRAM channel described by Section II.B, only the all-zero codeword is used for training. We also reduce the training set by just collecting the data around a given resistance spread level  $\sigma_{0} / \mu_{0}$. In this way the structure and time complexity of our proposed NNORB-MS decoder after training are similar to those of the regular RB-MS decoder, and the decoding also only involves integer/logic operations.

\section{Simulation Results}
The system parameters for our simulations are taken from \cite{zhang2011stt}, with $\mu_{0} = 1k\Omega$ and $\mu_{1} = 2k\Omega$. We further
assume $\sigma_{0} / \mu_{0}=\sigma_{1} / \mu_{1}$ according to the features of the memory fabrication process. We vary the resistance spread $\sigma_{0} / \mu_{0}$ (and hence $\sigma_{1} / \mu_{1}$), the mean $\mu_{b}$, and the normalized spread $\sigma_{b}/\mu_{1}$ of the unknown channel offset, to account for the influence of different process variations and the change of temperature. In our experiments, we implement and train the NNORB-MS algorithm by using the Tensorflow library and an NVIDIA Tesla K40c GPU. Moreover, the (71,64) Hamming code \cite{mram} is used to validate our proposed decoder.





\begin{figure}[t]
\centering
\includegraphics[height=0.55\columnwidth]{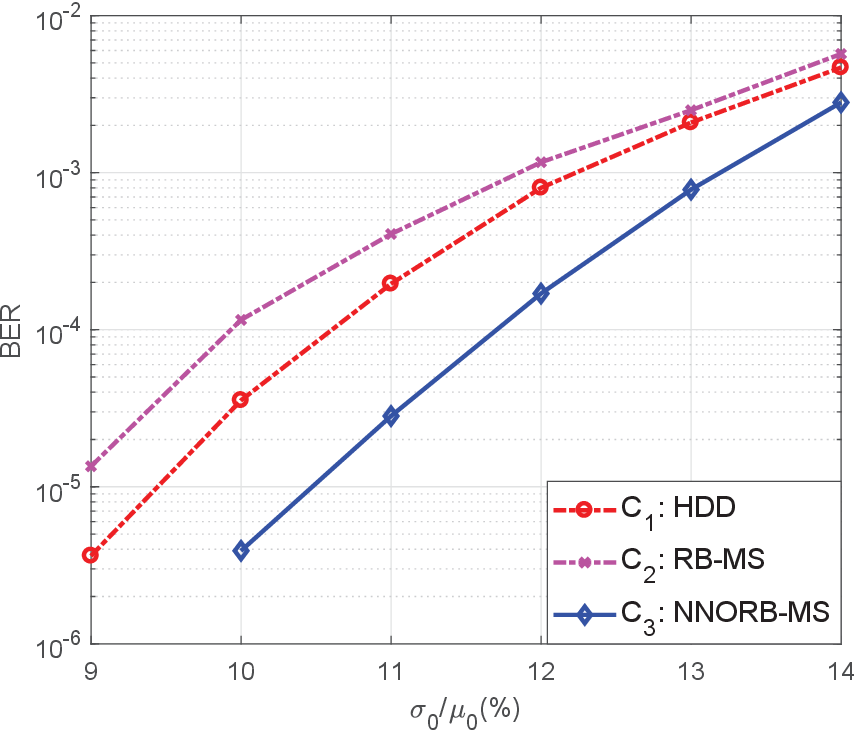}
\caption{BERs of different decoders for the channel without offset.}
\label{simulation_1}
\end{figure}

For the case without channel offset, Fig. \ref{simulation_1} shows that at bit error rate (BER) of $1\times10^{-5}$, our  {proposed} NNORB-MS decoder ($C_{3}$) outperforms the HDD ($C_{1}$) and the regular RB-MS decoder ($C_{2}$) by by $1.0\%$ and $1.7\%$, respectively, in terms of the maximum affordable resistance spread $\sigma_{0} / \mu_{0}$.

\begin{figure}[t]
\centering
\includegraphics[height=0.71\columnwidth]{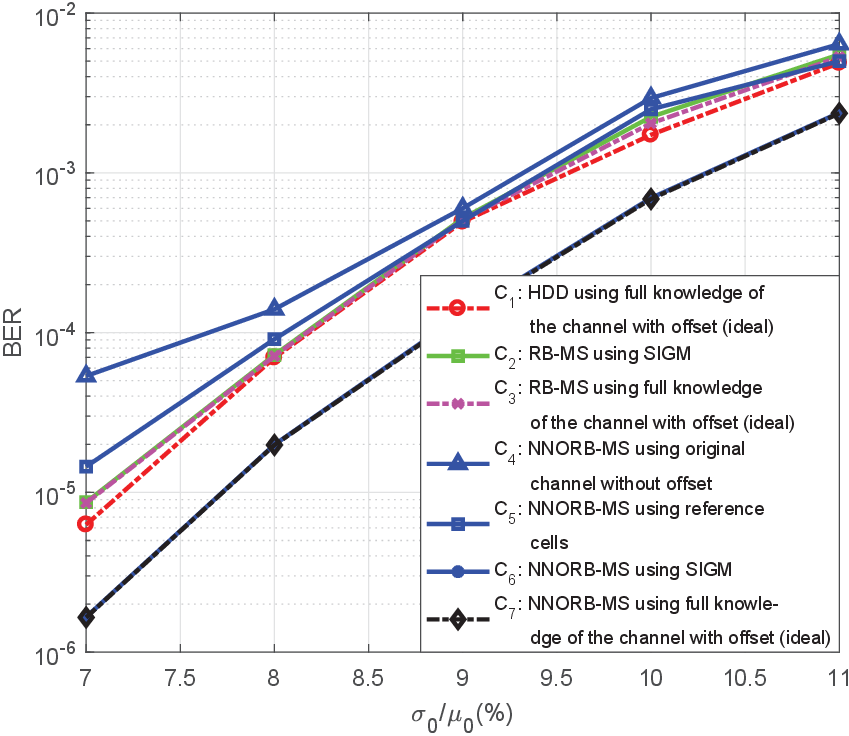}
\caption{BERs of different decoders for the channel with an offset of $\mu_{b} = -0.2 k\Omega$ and  $\sigma_{b} / \mu_{1}  = 4 \%$.}
\label{simulation_2}
\end{figure}

\begin{figure}[t]
\centering
\includegraphics[height=0.71\columnwidth]{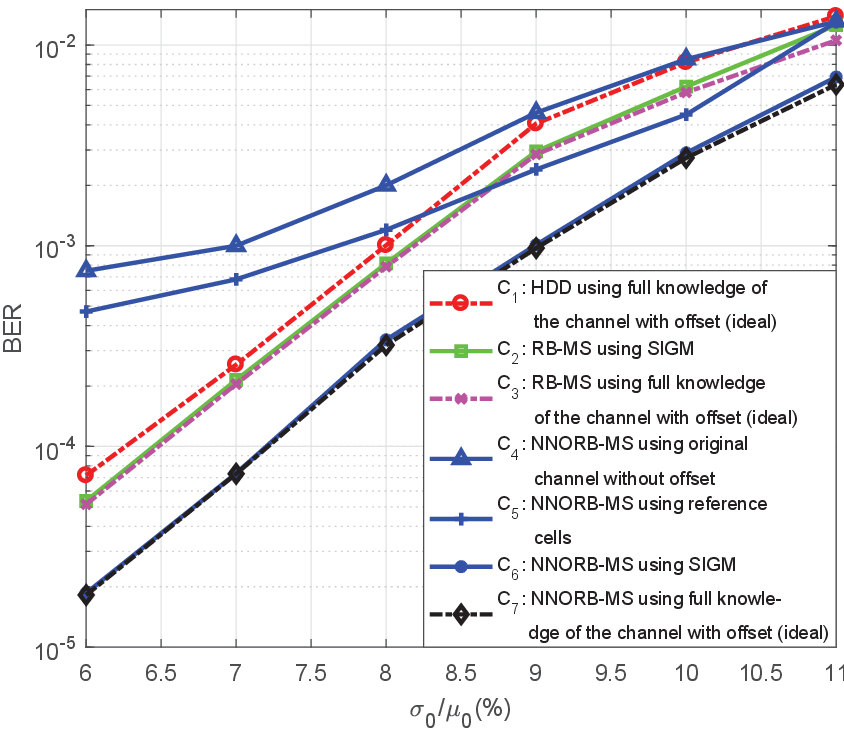}
\caption{BERs of different decoders for the channel with an offset of $\mu_{b} = -0.2 k\Omega$ and  $\sigma_{b} / \mu_{1}  = 7 \%$.}
\label{simulation_3}
\end{figure}

Next, we investigate the cases of the channel with  {unknown} offset. Fig. \ref{simulation_2} and Fig. \ref{simulation_3} illustrate the performance that the channel has an offset with a fixed mean of $\mu_{b} = -0.2k\Omega$, and different variations of $\sigma_{b} / \mu_{1}$ of $4 \%$ and $7 \%$, respectively. In both figures, we observe that the BERs of  {$C_{4}$} is poor even with the NNORB-MS decoder, if the decoder is trained and tested using the soft information without taking into account the channel offset ({\it i.e.} the decoder only knows the original channel without offset). By using the reference cell \cite{na2014reference} scheme, the approximate mean value $\mu_{b}$ of the channel offset can be obtained from the readback signals of the reference cells and used in the calculation of channel soft information. However, both figures show that the BER improvement with the reference cell scheme ({\it i.e.}  {$C_{5}$}) is very limited since the variance of the channel offset has not been taken into account. On the other hand, by using our proposed NN-based SIGM, accurate channel soft information can be generated. Fig. \ref{simulation_2} and Fig. \ref{simulation_3} show that the NNORB-MS decoder trained and tested using the SIGM ( {$C_{6}$}) significantly outperform  {$C_{4}$} and  {$C_{5}$}, and approaches the performance of  {$C_{7}$} where the decoder is trained and tested by assuming the full knowledge of the channel with offset is available (which is an ideal case and not practical). For this  {ideal} case, we derive the corresponding channel soft information by using the quantization scheme presented in \cite{mei2018information}.  {The NNORB-MS decoder} also provides an obvious BER gain over the HDD ($C_{1}$) and the regular RB-MS decoder ( {$C_{3}$}), whose inputs are also generated for the ideal case that the full knowledge of the channel is available.  {The NNORB-MS decoder outperforms the RB-MS decoder using the soft information generated by the NN-based SIGM ($C_{2}$) as well.} This demonstrates {thoroughly} the potential of our proposed NNORB-MS decoding for improving the reliability of STT-MRAM.

\section{Conclusions}
 We have developed novel DL-based algorithms and techniques to improve the decoding performance of linear block codes for STT-MRAM. Since the STT-MRAM channel is asymmetric, we have first symmetrized the channel using an i.i.d. channel adapter so that only the all-zero codeword is needed for training the NN decoder. In view of the unknown offset of the channel  {caused by the temperature change,} we have presented a novel NN-based SIGM to estimate the channel soft information without any prior knowledge of the channel. We have further proposed a novel NNORB-MS algorithm for decoding short linear block codes. We have demonstrated that it can significantly outperform both the HDD and the regular RB-MS algorithms over the STT-MRAM channel, with similar decoder structure and time complexity of the regular RB-MS decoder.

\section*{Acknowledgement}
This work is supported by Singapore Ministry of Education Academic Research Fund Tier 2 MOE2016-T2-2-054 and RIE2020 Advanced Manufacturing and Engineering (AME) programmatic grant A18A6b0057.

\small
\bibliographystyle{IEEEtran}
\bibliography{postdoc_refs}





\end{document}